\documentclass{jpsj-suppl}
\usepackage{txfonts} %Please comment out this line unless the txfonts package is availabe in your LaTeX system.

\title{The ${\bar{\rm P}}$ANDA Experiment at FAIR - Subatomic Physics with Antiprotons}

\author{Johan \textsc{Messchendorp}$^{1}$ on behalf of the ${\bar{\rm P}}$ANDA Collaboration}

\inst{$^{1}$KVI-CART/University of Groningen, Zernikelaan 25, 9747 AA Groningen, The Netherlands}

\email{j.g.messchendorp@rug.nl}

\recdate{September 16, 2016}

\abst{The non-perturbative nature of the strong interaction leads to spectacular phenomena, such as the formation
of hadronic matter, color confinement, and the generation of the mass of visible matter. 
To get deeper insight into the underlying mechanisms remains one of the most challenging tasks within the field
of subatomic physics. The antiProton ANnihilations at DArmstadt (${\bar{\rm P}}$ANDA) collaboration has the ambition to 
address key questions in this field by exploiting a cooled beam of antiprotons at the High Energy Storage Ring (HESR) 
at the future Facility for Antiproton and Ion Research (FAIR) combined with a state-of-the-art and versatile detector. 
This contribution will address some of the unique features of ${\bar{\rm P}}$ANDA that give rise to a promising physics
program together with state-of-the-art technological developments.}

\kword{antiprotons, hadrons, new facilities}

\begin{document}
\maketitle

\section{Introduction}

\begin{figure}[tbh]
\begin{center}
\includegraphics[width=0.9\textwidth]{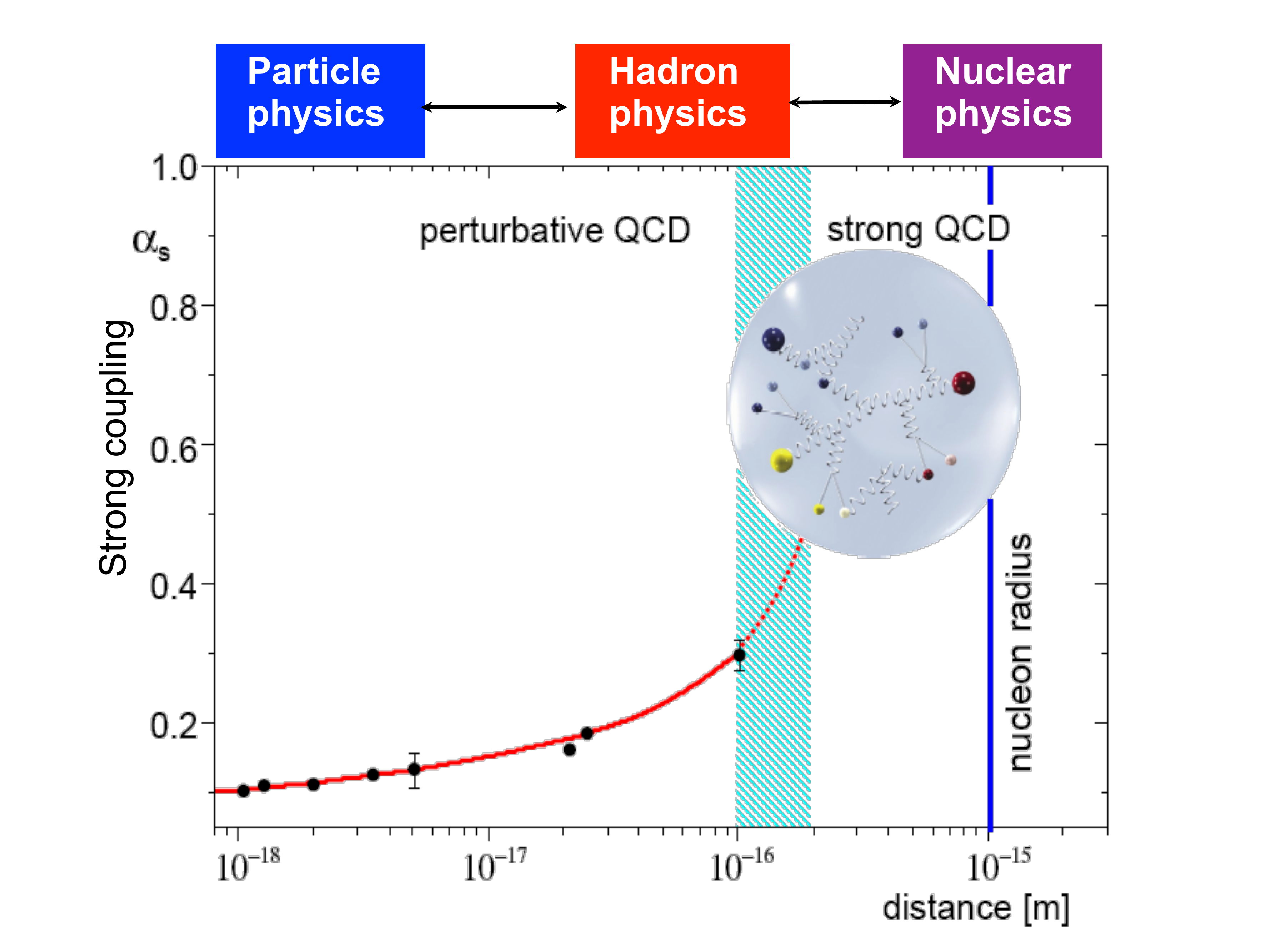}
\end{center}
\vspace*{-0.2cm}
\caption{The strong coupling constant as a function of distance scale. The ambition of ${\bar{\rm P}}$ANDA 
is to provide data that helps to bridge our understanding of QCD from short distance scales 
(quarks and gluons) to the scale of (hyper)nuclei (baryons and mesons).}
\label{strong-coupling}
\end{figure}

The physics of strong interactions is undoubtedly one of the most challenging areas of modern science.
Our present elegant and ``simple'' underlying fundaments of the strong interaction, Quantum Chromodynamics (QCD), is reproducing the 
physics phenomena only at distances much shorter than the size of the nucleon, 
where perturbation theory can be used yielding results of high precision and predictive power. At larger distance scales, 
however, perturbative methods cannot be applied anymore, although spectacular phenomena - such as the generation of hadron 
masses, the formation of hadronic matter, and color confinement - occur. It remains puzzling to identify the relevant degrees of freedom 
that connect the perturbative regime, driven by quarks and gluons, to the strong regime, which eventually leads to the 
formation of nuclei in which colorless pions and nucleons are the fundamental building blocks (see Fig.~\ref{strong-coupling}). 

The ${\bar{\rm P}}$ANDA (antiProton ANnihilation at DArmstadt)
collaboration will address various questions related to the strong 
interactions by employing a multi-purpose detector system~\cite{physperf} at the 
High Energy Storage Ring for antiprotons (HESR) of the upcoming Facility 
for Antiproton and Ion Research (FAIR). The ${\bar{\rm P}}$ANDA 
collaboration aims to connect the perturbative and the non-perturbative QCD regions, 
thereby providing insight in the mechanisms of mass generation and 
confinement. The collaboration is composed of an international spectrum of researchers 
from the nuclear, hadron, and particle physics communities, hence, covering the
complete field of subatomic physics.

The key ingredient for the ${\bar{\rm P}}$ANDA physics program is a high-intensity and 
a high-resolution beam of antiprotons in the momentum range of 1.5 to 15~GeV/$c$.
Such a beam gives access to a center-of-mass energy range from 2.2 to 5.5~GeV 
in $\bar p$$p$ annihilations. In this range, a rich spectrum of hadrons with
various quark configurations can be studied. In particular, hadronic states which contain 
charmed or strange quarks and gluon-rich matter are expected to be abundantly
produced in $\bar p$$p$ annihilations. 

The ${\bar{\rm P}}$ANDA detector will be installed at the HESR at the future FAIR facility.
Antiprotons will be transferred 
to the HESR where internal-target experiments 
can be performed using beams with unprecedented quality and intensity. 
Stochastic phase space cooling (and possibly electron cooling) 
will be available to allow for experiments with a momentum resolution in the range 
from $0.4-2\times 10^{-4}$ at luminosities up to $2\times 10^{32}$~cm$^{-2}$s$^{-1}$
corresponding to about $10^{11}$ stored antiprotons in the ring.
We note that the maximum number of stored antiprotons will be in the order of $10^{10}$ in the 
modularized start version of FAIR.

The usage of high-intense and cooled antiprotons covering a large center-of-mass range makes 
${\bar{\rm P}}$ANDA unique in the sense that it enables a {\it strange} and a {\it charm} ``factory'' 
providing a rich database for precision and exploratory studies. Such versatility also poses challenges in 
the development of the detector and its data processing. In the following, a more thorough discussion 
on these aspects will given, highlighting the magical aspects of antiprotons in terms of addressing
forefront physics questions and contributing to the next generation technology. 
 
\section{Versatility of antiprotons}

The antiprotons that will be used by ${\bar{\rm P}}$ANDA will enable a diverse physics program
with unique capabilities that cannot be found at other facilities. The following aspects
are at the basis of this uniqueness:

\begin{itemize}
 \item {\bf Large mass-scale coverage.} Center-of-mass energies from 2 to 5.5 GeV/$c^2$ are accessible
with the antiproton-momentum range covered at HESR. This will give access to hadronic states made from
light, strange, and charm quarks. The various quark-mass scales covered by ${\bar{\rm P}}$ANDA give a remarkable opportunity 
to systematically study the underlying degrees of freedoms in hadrons. Moreover, unconventional hadronic states with a 
gluon-rich nature, or composed from more than three quarks, are expected to be present in this energy regime. 
The recent discoveries of X, Y, Z states clearly demonstrate that this regime is rich in unconventional states.
But also the large momentum range gives access to the associated production of pairs of hadrons and antihadrons
with open strange- or/and charmness, such as $\Lambda\bar{\Lambda}$ and $D_s\bar{D}_s$ pairs.

 \item {\bf High hadronic production rates.} In contrast to experimental facilities that make use of electromagnetic probes (such as $e^+e^-$ colliders,
$e/\mu$-scattering facilities, and tagged-photon beams), ${\bar{\rm P}}$ANDA exploits hadronic interactions in the initial state, 
thereby taking advantage of large production cross sections and a high sensitivity to gluon-rich matter. In essence, ${\bar{\rm P}}$ANDA
will be a charm and strange factory providing high statistics data.

 \item {\bf Access to large spectrum of $J^{PC}$ states.} The direct formation of meson-like states is feasible in $\bar{p}p$ annihilation,
similar to, for example, the abundant charmonium production in $e^+e^-$. This aspect is attractive since it allows to study
the basic properties, such as the natural width of very narrow states via the resonance scanning technique. 
Antiproton-proton annihilation has the advantage that it can form states with all conventional $J^{PC}$ spin parities, 
whereas in $e^+e^-$ annihilation vector spin-parity states in formation, $J^{PC}=1^{--}$, are produced. 
In the latter case, other spin-parity states can be produced only indirectly 
via the decay of a vector state. Even in the indirect case, higher spin states remain unreachable as a consequence of the 
angular-momentum barrier. Antiproton-proton annihilations would, however, be sensitive to high spin states as well.  

\end{itemize}

Driven by the above-mentioned features, ${\bar{\rm P}}$ANDA has formulated an extensive physics program addressing
the ``hot spots'' in hadron physics using spectroscopical techniques and reaction dynamics. Details, including 
the results of feasibility studies using Monte Carlo (MC) simulations, have been reported in Ref.~\cite{physperf}.
These physics goals are supported by theorist and representatives of currently running and
planned experiments in the field~\cite{lut2016}. 
To summarize, the following items have been identified (with a large mutual overlap):

\begin{itemize}
 \item {\bf Hadron spectroscopy and dynamics.} Spectroscopy of hadronic states has been extremely successful in
the past and has led to the development of QCD and the quark model for baryons and mesons. QCD also predicts new
forms of hadronic matter such as multiquark states, glueballs, and hybrids. The existence of these non-conventional
forms of matter has been recently discovered experimentally by the observation of the so-called
X, Y, Z states. Their internal structure, e.g. the relevant degrees of freedom that lead to their formation, remains a mystery. 
The annihilation of antiprotons with protons or neutrons will be a complementary approach to systematically 
study their behavior. In particular, the sensitivity of probing high spin states and the possibility to form directly
unconventional states with conventional quantum numbers are of great advantage that cannot be done elsewhere. 
With ${\bar{\rm P}}$ANDA it would, therefore, be possible to determine precisely basic properties such as the line shape 
of some of the very narrow states and expand our spectroscopical knowledge. One also expects that with antiprotons it
would be possible to abundantly produce unconventional hadronic states, in particularly gluon-rich matter. 

 \item {\bf Nucleon structure.} Since in the case of ${\bar{\rm P}}$ANDA, the initial state is composed of a proton and an antiproton, 
it is feasible to study the structure of the nucleon as well, and in a way that is complementary to the 
traditional electron-scattering technique.  ${\bar{\rm P}}$ANDA will exploit the electromagnetic probe, 
the exchange of a virtual photon, in the time-like region, a regime that is not well explored yet but theoretically
as important as the space-like region. ${\bar{\rm P}}$ANDA will access time-like electromagnetic form factors with 
unprecedented accuracy via the annihilation of the antiproton with the proton and an 
electron-positron or negative-positive muon pair in the final state~\cite{sin2016}. In addition, transition distribution amplitudes (TDA) via
meson production, generalized distributed amplitudes (GDA) via hard exclusive processes and Compton scattering, 
and transverse parton distribution functions (TPD) via Drell-Yan production are foreseen to be studied by ${\bar{\rm P}}$ANDA.

 \item {\bf Hyperons and hypernuclei.} Baryons with strangeness extend the study of nucleon structure to the SU(3) domain. 
The additional strange quark probes a different QCD scale and, thereby, helps to systematically study the dynamics of
three-quark systems. The self-analyzing property of their weak decays provides a new rich of observables from which
one gains more insight in the spin properties of baryons. Moreover, a good understanding of the interaction among baryons 
with strangeness, e.g. on the level SU(3), is of crucial importance in view of understanding the properties of astrophysical 
objects such as neutron stars. At present, our incomplete understanding of the underlying meson-baryon, baryon-baryon and
multi-body interactions in baryonic systems limits probably our knowledge of the flavor composition of neutron stars. 
${\bar{\rm P}}$ANDA will be able to produce copiously hyperon pairs in antiproton-proton annihilations, acting as 
a strangeness factory. This will provide the basis to carry out an extensive baryon structure program. In a dedicated setup with a 
secondary target in combination with an additional array of Germanium gamma-ray detectors, it is foreseen to 
study the radiative level scheme of hypernuclei. The available antiproton momentum will, furthermore, allow to
produce pairs of open-charm baryons and mesons. This will extend the baryon and meson spectroscopy and production dynamics
with the much heavier charm quark as scale. 

 \item {\bf Hadrons in nuclear medium.} ${\bar{\rm P}}$ANDA plans to use heavy nuclear targets as well, thereby, giving
access to the study of antiproton-nucleus collisions. One of the physics interests lies in studying the properties
of hadrons, produced in the antiproton interaction, inside the nuclear environment. At finite nuclear density one expects 
that chiral symmetry is partially restored leading to a modification of the masses of hadrons. This would provide
the opportunity to study systematically the mass-generation mechanism. Another aspect that is on the ${\bar{\rm P}}$ANDA physics
with nuclear targets is the usage of the nucleus as a laboratory to determine the distance and time scale of hard QCD reactions.
This relates to the phenomenon known as ``Color Transparency (CT)''. Last, not least, is to study the short-distance
structure of the nucleus itself. Hard collisions of antiprotons inside the nucleus can be used as a probe to study
``Short Range Correlations (SRC)'', e.g. correlation effects of high momentum nucleons and non-nucleonic components of
the nuclear wave function.  

\end{itemize}

In the following sections, a few physics topics will be discussed of the above list that serve as key examples illustrating
the uniqueness and competitiveness of ${\bar{\rm P}}$ANDA. Note that the choice is partly driven by personal interest.
Moreover, topics that are potentially suited for a ``day one'' program with a somewhat limited intensity will be highlighted.
   
\section{Discovery by precision and exploration}

${\bar{\rm P}}$ANDA will cover center-of-mass energies up to 5.5~GeV/$c^2$ 
in $\bar{p}p$ collisions, sufficiently high to extensively cover
open- and hidden-charm states. The level scheme of lower-lying 
bound $\bar c$$c$ states, charmonium, 
is very similar to that of positronium. These charmonium states
can be described fairly well in terms of heavy-quark potential models.
Precision measurements of the mass and width of the charmonium spectrum
give, therefore, access to the confinement potential in QCD.  
Extensive measurements of the masses and widths of the 1$^{--}$ $\Psi$ states 
have been performed at $e^+$$e^-$ machines where they can be formed directly
via a virtual-photon exchange. Other states, which do not carry the
same quantum number as the photon, cannot be populated directly, but only
via indirect production mechanisms. This is in contrast to the $\bar p$$p$ 
reaction, which can form directly excited charmonium states of 
all conventional quantum numbers. As a result, the resolution in the mass and 
width, or more generally the line shape, of charmonium states can be determined by the 
precision of the phase-space cooled beam momentum distribution and not by the (significantly poorer) 
detector resolution. Moreover, in $\bar p$$p$ annihilation, charmonium states with high spin contents can
be populated directly, which cannot be done in $e^+e^-$ annihilation or in decays of
higher-lying states. Hence, ${\bar{\rm P}}$ANDA will be able to expand the existing and near-future 
experimental activities in charmonium(-like) research performed by BESIII, LHCb, BelleII, etc.. 
The need for such a tool becomes evident by 
reviewing the many open questions in the charmonium sector.
For example, our understanding of the states above the $D$$\bar D$ threshold 
is very poor and needs to be explored in more detail. There are various charmonium
states predicted, but yet to be discovered. Even more strikingly, there are many
states experimentally found that cannot be assigned to a conventional charmonium state.
These so-called X, Y, Z states got recently lots of attention. In particular, the charged
Z states gave rise to an excitement in the field, since their first-time confirmed discovery proofs unambiguously
that hadronic states composed of at least four quarks do exist. Even more recently, and in the
same energy interval, signals that strongly hints towards pentaquarks were found. 
Both discoveries of Z-states and pentaquarks were labeled as physics highlights by the 
American Physical Society (APS) in 2013 and 2015, respectively, demonstrating the
high impact of this type of research. Refs.~\cite{physperf,lan2013,pren2014} describe
the feasibilities of ${\bar{\rm P}}$ANDA for this physics topic in more detail.

To illustrate the potential of ${\bar{\rm P}}$ANDA in the field of X, Y, Z spectroscopy,
consider the X(3872). This state, discovered more than 10 years ago by Belle and confirmed
by many other experiments, has a natural width of less than 1.2~MeV (90\% confidence level), 
lies suspiciously close
to the $DD^*$ threshold, has a large isospin breaking, and has a spin-parity of $1^{++}$.
Based on these properties, it is expected that it is not a conventional $c\bar{c}$ configuration
but a candidate for a four-quark state, a $DD^*$ molecule, cusp effect, or any other
exotic combination involving charm-anticharm quarks. It is an experimental challenge to 
provide observables that would shed light on its true nature. One of the most 
sensitive parameters that would depend strongly on its nature, would be the actual lineshape or
natural width of the state~\cite{han2007}. The resonance scanning technique offered by using cooled antiprotons
would be ideal to study the lineshape of this very narrow state. Detailed MC simulations
were carried out to demonstrate the feasibility to use the observation of the channel 
$\bar{p}p\rightarrow {\rm X}(3872) \rightarrow J/\psi\pi^+\pi^-$ with a resonance scan to study the lineshape.
The simulation was based on a full detector model~\cite{spar11} with realistic background conditions, 
based on the Dual Parton Model~\cite{dpm} with a total $\bar{p}p$ cross section of 46~mb and a 
non-resonant $\bar{p}p\rightarrow J/\psi\pi^+\pi^-$ contribution of 1.2~nb~\cite{chen08}, a presumed production cross 
section for $\bar{p}p\rightarrow X(3872)$ of 100~nb, and a branching fraction of 5\%
for the decay X(3872)$\rightarrow J/\psi\pi^+\pi^-$. With the HESR ``day one'' condition of a 
luminosity of 1170~(nb$\cdot$day)$^{-1}$ and a center-of-mass energy resolution of 84~keV, 
a natural width for a Breit-Wigner response
of 100~keV can be measured with a precision of about 20\% in about 40 days. Such a sensitivity would
only be possible with an experiment like ${\bar{\rm P}}$ANDA.

In the discussion above, an example was given on how ${\bar{\rm P}}$ANDA can make an impact by
exploiting the excellent resolution of the antiproton beam. Another 
advantage of ${\bar{\rm P}}$ANDA is its high hadronic interaction rates which provide the basis
to carry out precision and exploration studies by statistics. ${\bar{\rm P}}$ANDA will be in essence
a factory of strange and charm hadrons. The hyperon sector is one of the topics that illustrate nicely the
advantages of ${\bar{\rm P}}$ANDA in this respect. These studies are largely motivated by 
extending our knowledge in the SU(2) sector to SU(3), e.g. what happens if we replace one of the
lighgt quarks in the proton with one - or many - heavier quark(s)? Moreover, the dynamics in which
hyperons are produced in hadronic interactions are of great interest, since it helps to shed light
on the degrees of freedom that are relevant at different mass scales. 
Already past experiments using antiprotons at LEAR gave a rich database in the associate 
production $\bar{p}p\rightarrow \bar{\Lambda}\Lambda$ near its production threshold. The database
above a beam momentum of 4~GeV/$c$ remains scarce. Furthermore, only a few bubble chamber events 
have been observed for $\bar{p}p\rightarrow \bar{\Xi}\Xi$, and no data exists for systems with 
an absolute strangeness of three ($|S|$=3), e.g. $\bar{p}p\rightarrow \bar{\Omega}\Omega$, nor in
charmed hyperons, $\bar{p}p\rightarrow \bar{\Lambda}_c\Lambda_c$. With ${\bar{\rm P}}$ANDA it would
become possible to produce strange hyperon pairs with typical observed rates, exclusively measured
using favorable decay modes, ranging from a few per hour ($\Omega^+\Omega^-$) to tens per second 
($\bar{\Lambda}\Lambda$, $\bar{\Lambda}\Xi^0$) at a ``day-one'' luminosity of 10$^{31}$~cm$^{-2}$s$^{-1}$.  
With these high rates, it would even become possible to obtain for the first time a rich set of polarisation 
observables by making use of the self-analyzing feature of weak decays. 

The high production rate of hyperon pairs in antiproton collisions is the basis for 
the novel hypernuclear program of ${\bar{\rm P}}$ANDA.
Details of the plans and feasibility of the hypernuclei program can be found in Refs.~\cite{lor12,sin16}. 
 The goal is to expand the nuclear chart
in the dimension of strangeness, thereby providing data that would help us studying two- and many-body
baryon-baryon forces with the inclusion of strangeness. ${\bar{\rm P}}$ANDA focusses on strangeness
$S=-2$ systems, $\Xi$-atoms and $\Lambda\Lambda$ hypernuclei. The hypernuclei program of ${\bar{\rm P}}$ANDA 
complements the research that is ongoing or planned at facilities like J-PARC~\cite{kan-jparc}, 
STAR~\cite{starhyp}, ALICE~\cite{alicehyp}, CBM and NUSTAR~\cite{fairhyp}, by 
providing the environment that allows to study $\gamma$-unstable excited states of double-hypernuclei.
To produce these hypernuclei, a primary diamond filament target will be bombarded by the antiproton beam, 
producing $\Xi^-$ baryons in $\bar{p}N\rightarrow \Xi^-\bar{\Xi}$. The $\Xi^-$ particles will be
decelerated in a secondary target, for example a boron absorber with active silicon layers.
The $\Xi^-$ hyperons might be captured in an electron shell of an atom, forming hyperatoms. An
excited $\Lambda\Lambda$-nucleus can be formed via the conversion reaction $\Xi^- p \rightarrow \Lambda\Lambda$.
It will decay to its groundstate by the emission of $\gamma$-rays, which are detected by an array of Germanium
detectors. The pionic decays of the $\Lambda$ hyperons can be exploited to provide a clean signal.
Already at an antiproton interaction rate of 2$\times$10$^6$~s$^{-1}$, ${\bar{\rm P}}$ANDA will 
be able to reach a rate of stopped $\Xi^-$ which is comparable to the maximum rate expected at~\cite{kan-jparc}.

The formation of $\Xi^-$ atoms will be the first step towards a forefront game changer, namely a study of the 
spectroscopic quadrupole moment of the $\Omega^-$ via the hyperfine splittings in $\Omega^-$-atoms. 
The long lifetime and its spin of 3/2 makes the $\Omega^-$ the only candidate to obtain direct experimental information
on the shape of an individual baryon. Such a measurement would complement the world-wide studies
of the shape of the proton, with the interesting feature of addressing a $|S|=3$ system in which
meson cloud corrections to the valence quark core are expected to be small. The quadrupole moment
of the $\Omega^-$ would also serve as an ideal benchmark for lattice QCD calculations since the contributions 
from light quarks are small. Unfortunately, it is yet unknown what the production rates of $\Omega$ hyperons are
in antiproton collisions.

\section{Technological innovation}

\begin{figure}[tbh]
\begin{center}
\includegraphics[width=\textwidth]{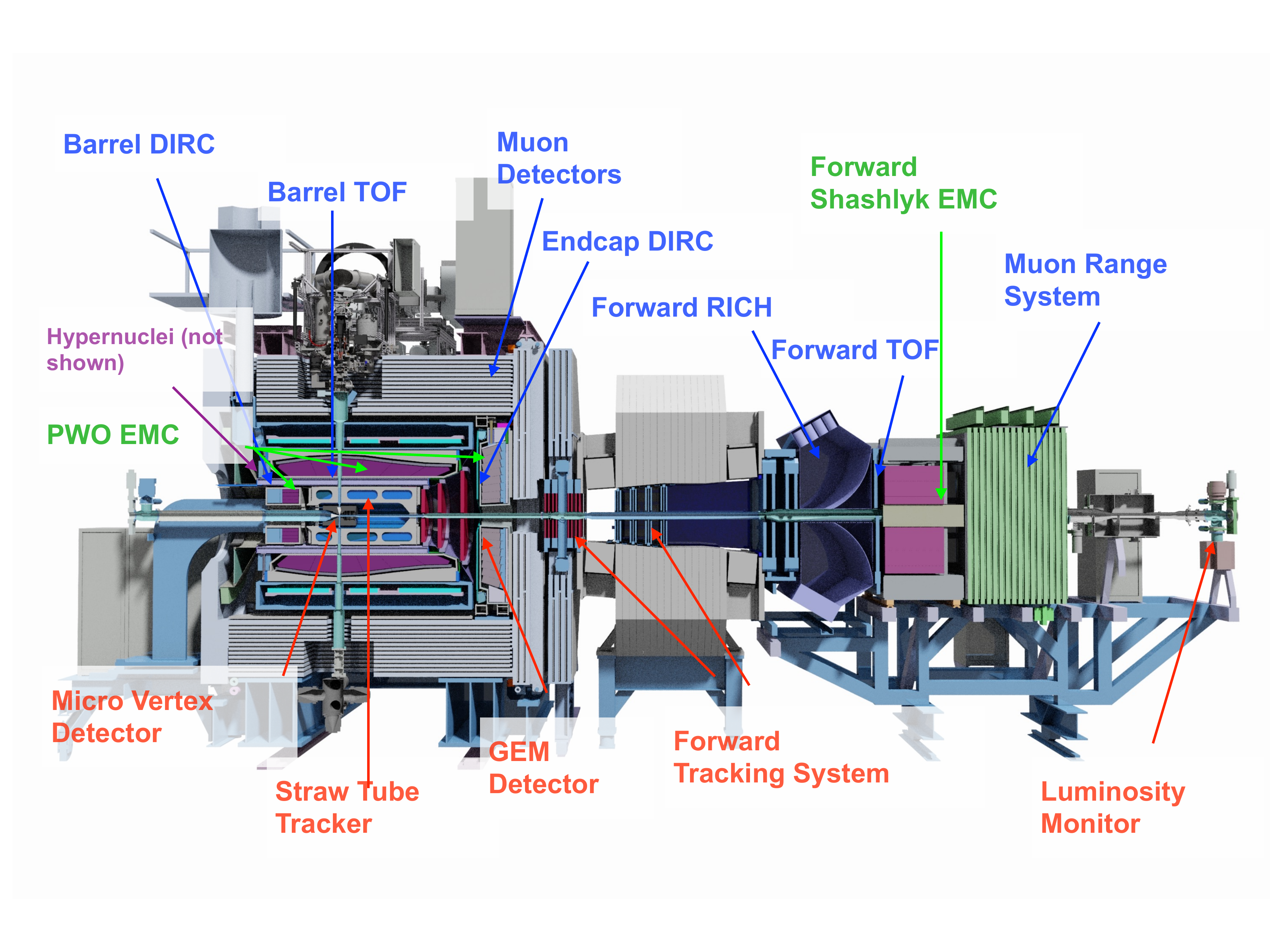}
\end{center}
\vspace*{-1cm}
\caption{A sketch and cross section of the full ${\bar{\rm P}}$ANDA setup. The various vertex, tracking, 
particle identification, and photon calorimeter components are indicated. More details can be found in Refs.~\cite{physperf,tdrs}.}
\label{panda-detector}
\end{figure}

The physics ambitions of ${\bar{\rm P}}$ANDA go hand-in-hand with technological 
detector developments. 
The $\bar{p}p$ cross section with the momentum range covered by ${\bar{\rm P}}$ANDA,
range from 50-100~mb. Such large cross sections together with an intense antiproton beam result
in enormous interaction rates going up to 2$\times$10$^7$~s$^{-1}$. 
In contrast, the ${\bar{\rm P}}$ANDA detector will be able to identify reactions with
cross sections of only a few nb, i.e. needle-in-a-haystack capabilities. 
To cope with the high rates, a highly granular
detector will be used with components that are capable of coping with
high count rates. To address the different physics topics, 
the detector needs to cope with a variety of final states and a large range of 
particle momenta and emission angles. A 4$\pi$ detection system is foreseen which is
necessary in order to unambiguously carry out a partial-wave decomposition. 
The broad physics program, that asks for hadronic, electromagnetic and weak probes, 
requires, furthermore, a 
detector that has excellent particle identification capabilities and a high momentum
resolution to be able to identify and measure the momentum of photons, electrons/positrons, 
muons, pions, kaons, and (anti)protons. The lead-tungstate photon calorimeter of 
${\bar{\rm P}}$ANDA will have a huge dynamic-range capability to detect photons 
from a few MeV to a few GeV in energy. To identify weakly decaying open-strangeness/charmness 
hadrons, ${\bar{\rm P}}$ANDA will be equipped with the capability to identify displaced vertices.
The experiment will use internal targets. It is conceived to use either pellets 
of frozen H$_2$ or cluster jet targets for the $\bar pp$ reactions, 
and wire targets for the $\bar pA$ reactions. Moreover, the setup has been designed
in a modular way, that would allow to easily replace the micro vertex detector and
the backward part of the electromagnetic calorimeter by the hypernuclear setup with
its dedicated targets and gamma-ray detectors.
A sketch of the ${\bar{\rm P}}$ANDA setup is depicted in Fig.~\ref{panda-detector} indicating the
various components that are foreseen to reach all the previously mentioned requirements.
More details can be found in various technical design reports~\cite{tdrs}.

For the full design of the detector as shown in Fig.~\ref{panda-detector}, an 
interaction rate of 2$\times$10$^7$~s$^{-1}$ would translate in a raw data rate of the order  
200~GBytes/s. It would be impossible and unpractical to simply store all data on disk or tape.
Since a conventional hardware trigger will not be possible, a new paradigm in data 
processing is being developed. The aim is to deploy a free-streaming data-processing scheme, 
whereby the challenge 
lies in reconstructing {\it in-situ} the complete event topology for each $\bar{p}p$ 
interaction under high-countrate conditions and using massive parallization hardware
architectures. Feasibility studies show promising results~\cite{kav12,gal13,adi2014,Stoc2015,bab2015}.   

\section{Summary}

The ${\bar{\rm P}}$ANDA experiment at FAIR will address a wide range of
topics in the field of QCD, of which only a small
part could be highlighted in this paper. The physics program
will be conducted by using beams of antiprotons together with 
a multi-purpose detection system, which enables
experiments with high luminosities and precision resolution.
${\bar{\rm P}}$ANDA has the ambition to provide valuable and new insights in 
the field of hadron physics which would bridge 
our present knowledge obtained in the field of perturbative QCD 
with that of non-perturbative QCD and nuclear structure.
It offers a long-term program for the
next generation (nuclear, hadron, and particle) physicists.
Already at the starting phase of HESR, with a reduced
luminosity, various key experiments can be carried out, such
as a resonance scan of the X(3872), a search for high-spin states in
charmonium-like systems, a high-statistics 
study of hyperons and their production in $\bar{p}p$ annihilation, 
and many more. 

\section{Acknowledgments}

I thank Paola Gianotti, Maria Carmen, and Alfons Khoukaz for carefully reading
this proceeding and for tutoring the corresponding presentation given at MENU2016.
Moreover, I thank the organizers of MENU2016 for hosting an excellent conference
with many inspiring talks and discussions.

\end{document}